\documentclass[reprint,superscriptaddress]{revtex4-1}

\usepackage{amsmath}
\usepackage{amssymb}
\usepackage{color}
\usepackage{siunitx}
\usepackage{graphicx}
\usepackage{dcolumn}
\usepackage{bm}
\usepackage{hyperref}

\usepackage{booktabs}
\usepackage[table,xcdraw]{xcolor}

\begin{document}
\title{\textit{Ab initio} electron-phonon coupling theory of elastic helium atom scattering}
\author{Cristóbal Méndez}
\affiliation{School of Applied and Engineering Physics, Cornell University, Ithaca, NY, 14853}
\author{C. J. Thompson, M. F. Van Duinen, S.J. Sibener}
\affiliation{The James Franck Institute and Department of Chemistry, The University of Chicago, Chicago, Illinois  60637}
\author{Tom\'as A. Arias}
\affiliation{Department of Physics, Cornell University, Ithaca, New York 14853}

\date{\today}

\begin{abstract}
We propose a fully \emph{ab initio} approach to predicting thermal attenuation in elastic helium atom scattering amplitudes, validated through strong agreement with experiments on Nb(100) and (3$\times$1)-O/Nb(100) surfaces. Our results reveal the relative contributions from bulk, resonant, and surface phonon modes, as well as from different surface mode polarizations, providing insights into differences between smooth and corrugated surfaces. These findings advance understanding of surface dynamics and electron-phonon coupling, laying groundwork for future studies on surface superconductivity.

\end{abstract}

\maketitle

Helium atom scattering (HAS) has emerged as a powerful and versatile technique for probing the structural and dynamical properties of surfaces. HAS leverages the interaction between low-energy helium atoms and surface atoms, enabling detailed, non-destructive measurements of delicate surfaces\cite{holst2021material,tamtogl2021atom}. In HAS, incoming helium atoms primarily interact with the surface electron density, indirectly encoding valuable information about surface electron-phonon coupling, a topic of significant recent interest\cite{luo1993electron, manson2016electron}.

Due to the indirect relationship between helium atom scattering and electron-phonon coupling, interpreting experimental measurements requires reliable \emph{ab initio} methods that directly and explicitly treat electron-atom interactions. Recent studies have successfully described \emph{inelastic} HAS scattering at a fully \emph{ab initio} level for simple surfaces\cite{kelley2024fully}. However, inelastic scattering primarily detects individual phonons, which limits its ability to provide a comprehensive global view of electron-phonon coupling. Moreover, because this method focuses on single phonon events, extensive measurements are needed to gather sufficient statistics for determining electron-phonon coupling strengths across the Brillouin zone.

Elastic scattering, in contrast, directly averages across the Brillouin zone, offering a global view of electron-phonon scattering with robust detection statistics. A key feature of \emph{elastic} HAS is the attenuation of scattering intensity due to thermal effects. This phenomenon originates from random fluctuations in the surface electron density driven by thermally excited phonons near the surface. These fluctuations then reduce the coherence of reflected helium-atom wave functions, diminishing the final scattered intensities. This phenomenon mirrors the bulk Debye-Waller effect, where, for example, X-ray scattering peak intensities decrease exponentially with temperature according to the Debye-Waller factor \cite{benedek2020measuring, vila2007theoretical, benedek2014unveiling}.
Recent studies have shown that the Debye-Waller factor in helium atom scattering is directly proportional to the electron-phonon coupling constant, affording a novel way to study near-surface electron-phonon coupling in a variety of systems \cite{levi1979quantum,hulpke1992helium,manson2016electron}. Detailed measurements and analysis of Debye-Waller factors can thereby provide key insights into superconducting properties near surfaces and, potentially, of two-dimensional materials \cite{scalapino2018electron, gunst2016first}.

Previous theories of elastic HAS connect random out-of-plane atomic motions at the surface to the experimentally measurable Debye-Waller factor\cite{armand1977debye,thompson2024correlating,doi:10.1021/acs.jpcc.4c02430}. While some studies use \emph{ab initio} description of phonons, none are fully \emph{ab initio}, as they rely on arbitrary choices—such as focusing only on out-of-plane atomic motion at the surface and selecting the number of surface atoms to average over during collisions. The choices become more complex for multi-species systems, yielding a wide range of predictions for attenuation factors\cite{doi:10.1021/acs.jpcc.4c02430}. Moreover, these methods exclude electrons, fundamentally precluding extraction of insights into electron-phonon coupling from experimental data.

More promising is the approach proposed by Garc\'ia and Benedek\cite{garcia1976analysis,benedek1979theory}, which considers temperature-induced fluctuations in the ``hard corrugated surface'' (HCS), the locus of points from which incoming helium atoms can be considered to experience hard-wall collisions. However, Garc\'ia and Benedek's ground-breaking results require knowledge of the derivatives of the full two-dimensional hard corrugated surface with respect to the atomic positions, and have therefore eluded \emph{ab initio} application so far. Moreover, this approach does not directly involve electrons, again limiting the ability to extract electron-phonon coupling from experimental data.

\begin{figure*}[t]
   \centering
   \includegraphics[width=1.0\textwidth]{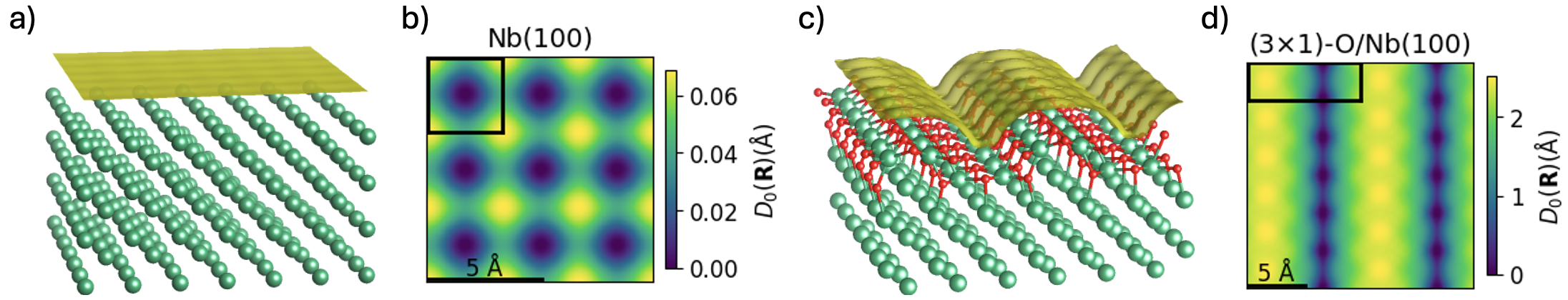}

\caption{Locus of turning points for elastic helium atom scattering from Nb(100) metal (a,b) and (3$\times$1)-O/Nb(100) oxide (c,d) surfaces for incident normal-component kinetic energies of $E_{i,z}$~=~14~meV and 12~meV, respectively: (a,c) niobium atoms (green spheres), oxygen atoms (red spheres), isodensity model turning points (gold surfaces); (b,d) height of isodensity model turning points $D_0(\mathbf{r})$ across the respective surfaces (blue-yellow color maps; note significant differences in scales).}   
   \label{fig:top}
\end{figure*}

The present work resolves the above issues by developing an \emph{ab initio} approach that explicitly addresses the electrons and thereby provides direct access to electron-phonon coupling. We first determine the hard corrugated surface from contours of the \emph{ab initio} surface electron density. Next, we develop a method to compute the necessary derivatives of the corrugated surface with respect to atomic positions. Finally, using the tangent-plane/Kirchhoff scattering approximation\cite{brekhovskikh1952difrakcya, brekhovskikh1952diffraction}, we directly link the temperature dependence of elastic helium atom scattering intensities to electron-phonon coupling. To confirm our results and derive new insights into the electronic and vibrational properties of Nb surfaces, we carry out the corresponding experiments and cross-correlate our theoretical and experimental results. Our fully \emph{ab initio} approach leads to predictions in excellent agreement with experiments for both the very mildly corrugated, single-species Nb(100) surface (Figs.~\ref{fig:top}\,a,b) and the highly corrugated, multi-species (3$\times$1)-O/Nb(100) oxidized surface (Figs.~\ref{fig:top}\,c,d). Additionally, our theory provides direct insight into the relative contributions of different phonon modes and offers \emph{ab initio} evidence supporting the proposed mechanism for the recently observed nonlinear behavior in elastic scattering attenuation exponents in certain systems\cite{benedek2022role}. 

\emph{Experimental procedure.---} All crystalline surfaces and helium atom scattering measurements were performed in our customized ultrahigh vacuum (UHV) scattering instrument, which has been described elsewhere\cite{mcmillan2020persistence, gans1992surface, niu1995phonons, mcmillan2022combined}. (Key details of the instrument are summarized in the Supplemental Materials.) For our experiments, we used He as our probe, scattering from a Nb(100) surface with a supersonic He atomic beam. Measurements were obtained for the thermal dependence of the intensity of Bragg peaks for the scattering of helium atoms at normal incident angle $\theta_i = 35^\circ$. We obtained data for an incoming kinetic energy $E_i = 21$~meV for the Nb(100) surface and at a slightly different energy of $E_i = 18$~meV for the (3$\times$1)-O/Nb(100) reconstructed oxidized surface. Note that, with the well-depth correction under the Beeby approximation \cite{scoles1988atomic}, these values correspond to effective normal incident energies $E_{i,z} = 14$~meV and $12$~meV, respectively.

\emph{Computational methods. ---} All \emph{ab initio} density-functional theory calculations employed our open-source plane-wave software JDFTx \cite{payne1992iterative,sundararaman2017jdftx}, with norm-conserving pseudopotentials \cite{schlipf2015optimization} and the Perdew-Burke-Ernzerhof functional revised for solids (PBEsol) \cite{perdew2008restoring}. (See supplemental materials for further details.) 

\emph{Ab initio} determination of hard corrugated surface. — In the hard corrugated surface (HCS) approach, elastic collisions of helium atoms are modeled as ``hard wall’’ interactions with a corrugated surface determined by the target's atomic positions\cite{garcia1976scattering, goodman1979model}. This surface is described mathematically by the \emph{corrugation function} $D(\mathbf{r})$, defined as the outward normal coordinate of the surface as a function of the two-dimensional in-plane vector $\mathbf{r}$---such that $(\mathbf{r},D(\mathbf{r}))$ gives coordinates on the corrugated surface \cite{benedek1983surface}.

To provide an \emph{ab initio} treatment of $D(\mathbf{r})$, we adapt a method from first-principles solvation studies, where the solvent-excluded cavity is defined by an isosurface of the solute electron density, $n(\mathbf{r}) = n_c$\cite{fattebert2003first, petrosyan2005joint}. Similar to solvated systems, the closed-shell nature of the helium atoms in HAS causes the nearby atom-surface interaction to arise primarily from Thomas-Fermi overlap of the respective electron clouds, which increases steeply with the surface electron density $n(\vec{r})$. This leads to nearly hard-wall interactions, as recently documented in \emph{ab initio} studies of helium atom surface collisions\cite{kelley2024fully}. We thus expect, as with solvated systems, that the hard-corrugated surface is well described by $n\left(\mathbf{r}, D(\mathbf{r})\right)=n_c$, where $n_c$ is chosen to yield scattering locations corresponding to the classical turning points of the incoming helium atoms.

To determine $n_c$, we compute the classical collision turning points from the \emph{ab initio} helium-surface interaction energy $E(\mathbf{r}, z)$. For each surface location $\mathbf{r}$, the vertical turning point $z_{KE}$ is identified where the potential equals the normal component of the incoming helium kinetic energy ($E_{i,z} = 14$~meV and $12$~meV for Nb(100) and (3$\times$1)-O/Nb(100), respectively). We then evaluate the \emph{ab initio} electron density of the non-interacting surface at these points, $n(\mathbf{r}, z_{KE})$, and determine $n_c$ from the resulting average value.

\begin{figure}[!htb]
   \centering
   \includegraphics*[width=1\columnwidth]{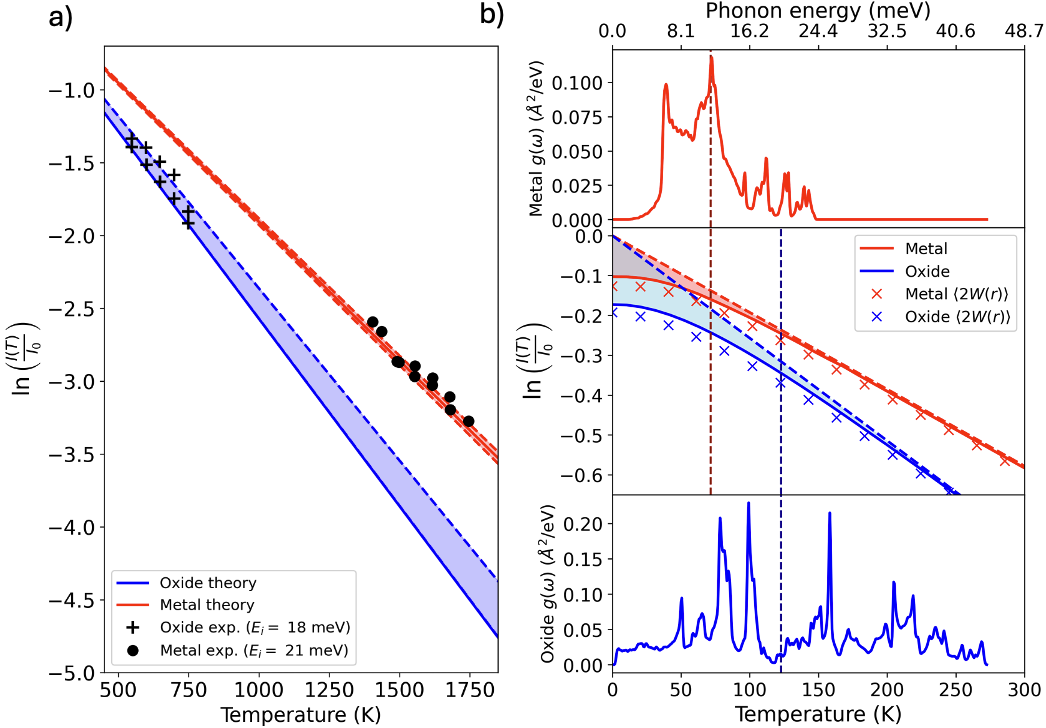}
   \caption{Logarithm of normalized specular intensity versus temperature for metal and oxide surfaces  for incoming helium atom energies of $E_i=21$~meV and $E_i=18$~meV, respectively, for (a) high- and (b) low-temperature regimes. (a) High-temperature regime: experimental data for Nb(100) (circles) and (3$\times$1)-O/Nb(100) (+'s), respective \emph{ab initio} predictions (red and blue ranges).
   (b) Low-temperature regime (\emph{ab initio} predictions only): normalized intensities (central panel) for metal (red curve) and oxide (blue curve) with linear extrapolations (red and blue dashed lines, respectively) with approximation from Eq.~\ref{eq:esa} (red and blue crosses). Upper and lower panels in (b) show spectral functions $g(\omega)$ for metal and oxide, respectively, with phonon energy scale adjusted to correspond to temperature in the central panel and vertical dashed lines indicating each spectral function's median value. (See text for details.)}
   \label{fig:intensity}
\end{figure}

For the smooth metallic Nb(100) surface (Figs.~\ref{fig:top}\,a,b), a uniform 5$\times$5 mesh across the primitive 1$\times$1 surface cell yields $n_c = 1.30 \pm 0.19$~e$^-$/nm$^3$. The relatively small variation in these computed $n(\mathbf{r},z_{KE})$ values,  combined with the small $\Delta z_{rms} = 0.07$~{\AA} root-mean-square difference between the heights of our computed turning-point hard corrugated surface and the heights of the density-contour surface (especially when compared to the de Broglie wavelength of $\sim 0.4$~{\AA} associated with the normal direction kinetic energy),  together support our proposed density-contour approach. Similarly, for the highly corrugated (3$\times$1)-O/Nb(100) oxide surface (Figs.~\ref{fig:top}\,c,d), a 15$\times$5 mesh yields $n_c = 0.66 \pm 0.20$e$^-$/nm$^3$, with $\Delta z_{rms} = 0.12$~{\AA}, much smaller than the surface's 2.8~{\AA} maximum corrugation. For both cases, in accord with the well-known \emph{anti-corrugation} effect \cite{senet2002theory, jean2004corrugating, petersen1996scattering}, we find the maximum discrepancy between the turning-point corrugated surfaces and our density contours to occur at the top sites, where the actual turning points for the Nb(100) and (3$\times$1)-O/Nb(100), respectively, are 0.15~{\AA} and 0.27~{\AA} \emph{deeper} than the density-contour surface. Regardless of any differences, for both cases, we find that our predictions below are quite insensitive to the choice of contour parameter $n_c$ across the above uncertainty ranges and are in excellent agreement with experiment (Fig.~\ref{fig:intensity}).

\emph{Ab initio} calculation of reflection amplitudes. — To compute reflection probabilities for the hard reflection surface described above, we utilize the tangent-plane/Kirchhoff approximation developed by Brekhovskikh\cite{brekhovskikh1952difrakcya, brekhovskikh1952diffraction, voronovich2013wave}, extending Isakovich’s statistical version\cite{isakovich1952scattering} to account for atomic fluctuations on periodic surfaces. This method is valid when $|\vec k_0| r_c \cos^3\theta \gg 1$, where $\vec k_0$ is the incoming wave vector, $r_c$ is the local surface curvature radius, and $\theta$ is the incidence angle relative to the surface normal\cite{lynch1970curvature}.  (Some authors list this condition with an even more favorable factor of $\cos\theta$ rather than $\cos^3\theta$; see  \cite{fung1981note} for further details.) For our density-contour determined reflection surface for Nb(100) in Figs.~\ref{fig:top}\,a,b, we find a root mean absolute Gaussian curvature of $r_c^{-1}=4.9 \times 10^{-4}$~\AA$^{-1}$, so that for our incoming kinetic energy and incident angle of $\theta \approx 35^\circ$, we find indeed that $|\vec k_0| r_c \cos^3\theta = 81.2 \gg 1$. For our much more corrugated (3$\times$1)-O/Nb(100) surface, we instead find $|\vec k_0| r_c \cos^3\theta = 5.7 > 1$, so that we can expect reasonable results here as well.

The Kirchhoff approximation then gives the scattering amplitude at wave vector $\vec{k}=(\mathbf{k},q)$ from an incident plane wave of wave vector $\vec{k}_0=(\mathbf{k}_0,-q_0)$ to be \cite{voronovich2013wave}
\begin{equation} \label{eq:S}
S(\vec{k},\vec{k}_0)=Q(\vec{k},\vec{k}_0) \int \frac{d^2\mathbf{r}}{(2\pi)^2} e^{-i(\mathbf{k}-\mathbf{k}_0)\cdot \mathbf{r}} e^{i(q+q_0)D(\mathbf{r})} ,
\end{equation}
where the kinematic prefactor is
\begin{equation}\nonumber
Q(\vec{k},\vec{k}_0) =-\frac{(q q_0)^{-\frac{1}{2}}}{q+q_0}\left(|\mathbf{k}|^2+q^2+q q_0 - \mathbf{k}\cdot \mathbf{k}_0\right),
\end{equation}
which has value -1 for specular reflection ($\mathbf{k}=\mathbf{k}_0$, $q=q_0$). Next, in the thermodynamic limit of surfaces much larger than the interatomic fluctuation correlation length, the expected intensity approaches $I\equiv \langle|S(\vec{k},\vec{k}_0)|^2\rangle \longrightarrow |\langle S(\vec{k},\vec{k_0})  \rangle|^2$. Finally, when computing $\langle S(\vec{k},\vec{k_0})  \rangle$ from Eq.~\ref{eq:S} averaged over all thermal configurations, the cumulant expansion naturally truncates at second order because $D(\mathbf{r})$ is Gaussian distributed\cite{messiah1967quantum}, guaranteeing that all higher-order cumulants are exactly zero regardless of the size of fluctuations in $D(\mathbf{r})$. The result is 
\begin{eqnarray} \label{eq:mS}
\langle S(\mathbf{k},\mathbf{k}_0)\rangle &=& Q(\mathbf{k},\mathbf{k}_0) \int \frac{d^2\mathbf{r}}{(2\pi)^2} e^{-i(\mathbf{k}-\mathbf{k}_0)\cdot \mathbf{r}}  \cdot\\ && \mbox{\ \ \ \ } e^{i(q+q_0)D_0(\mathbf{r})}  e^{-\frac{1}{2}(q+q_0)^2\delta D^2(\mathbf{r})}, \nonumber
\end{eqnarray}
where, echoing the notation of Garc\'ia and Benedek \cite{garcia1976analysis,benedek1979theory}, $D_0(\mathbf{r})$ and $\delta D^2(\mathbf{r})$ are the mean and variance, respectively, of the corrugation function $D(\mathbf{r})$ at in-plane location $\mathbf{r}$.

For periodic surfaces, both $D_0(\mathbf{r})$ and $\delta D^2(\mathbf{r})$ are periodic, enabling the reduction of the integral in Eq.~\ref{eq:mS} to an integral over the area $A$ of the surface unit cell,
\begin{eqnarray} \label{eq:pmS}
\langle S(\mathbf{k},\mathbf{k}_0)\rangle &=& Q(\mathbf{k},\mathbf{k}_0) \sum_{\mathbf{G}} \delta(\mathbf{k}-\mathbf{k}_0-\mathbf{G})  \cdot \\ &&  \int_A \frac{d^2r}{A} e^{-i \mathbf{G}\cdot \mathbf{r}}    e^{i(q+q_0)D_0(\mathbf{r})} e^{-\frac{1}{2}(q+q_0)^2\delta D^2(\mathbf{r})}, \nonumber
\end{eqnarray}
where $\mathbf{G}$ ranges over the surface reciprocal lattice. Here, the Dirac comb maintains the Bragg peaks, and the prefactor and integral give the strengths of those peaks. Benedek \emph{et al.} \cite{benedek1981inelastic,benedek1983surface} identify the real exponent
\begin{equation} \label{eq:WR}
W(\mathbf{r})=\frac{1}{2}\left(\Delta k_z\right)^2 \delta D^2(\mathbf{r}),
\end{equation}
where $\Delta k_z\equiv q+q_0$ is the normal momentum transfer to the surface, as the \emph{position-dependent Debye-Waller factor}. It is the specific integral over $A$ in Eq.~\ref{eq:pmS} that determines how these $\mathbf{r}$-dependent Debye-Waller factors must be combined to determine the temperature-dependent diminution of the Bragg peaks.

Using the electron density to define $D(\mathbf{r})$ enables a fully \emph{ab initio} determination of surface corrugation that directly links corrugation fluctuations to electron-phonon coupling. This connection arises explicitly through the derivatives $\partial_{a\alpha} n(\vec{r}) \equiv \partial n(\vec{r}) / \partial u_{a\alpha}$ of the target electron density with respect to the atomic displacements $u_{a\alpha}$ of each atom $a$, derivatives which are readily obtained from standard Wannier-based electron-phonon coupling computations\cite{marzari1997maximally, giustino2007electron}. Variational analysis for small atomic displacements of the condition $n(\vec{r}) = n_c$ gives $0 = \delta n = \delta D \cdot \partial_{\hat{n}} n + \sum_{a\alpha} u_{a\alpha} \partial_{a\alpha} n$, where $\partial_{\hat{n}} n$ is the surface-normal directional derivative of the electron density. This relationship allows the corrugation fluctuations to be extracted in terms of $\partial_{\hat{n}} n$ and $\partial_{a\alpha} n$,
\begin{equation} \label{eq:dD2}
\delta D^2(\mathbf{r})  = \sum_{a,b,\alpha,\beta} \left\langle u_{a \alpha} u_{b \beta}\right\rangle \frac{\partial_{a\alpha}n(\mathbf{r},D(\mathbf{r}))}{\partial_{\hat n} n(\mathbf{r},D(\mathbf{r}))} \frac{\partial_{b\beta}n(\mathbf{r},D(\mathbf{r}))}{\partial_{\hat n} n(\mathbf{r},D(\mathbf{r}))}.
\end{equation}
Here, the usual phonon-correlation function $\langle u_{a \alpha} u_{b \beta}\rangle$ is available fully \emph{ab initio},
\begin{eqnarray} \label{eq:uu}
\left\langle u_{a \alpha} u_{b \beta}\right\rangle&=&\frac{\hbar}{2 \sqrt{m_a m_b}}\cdot\\ && \left\langle \sum_{j} \frac{1+2 N_{\mathbf{q} j}}{\omega_{\mathbf{q} j}}\hat e_{\mathbf{q} j}^{* a \alpha}\hat e_{\mathbf{q} j}^{b \beta} e^{i\mathbf{q}\cdot (\mathbf{x_b}-\mathbf{x_a})}\right\rangle_{\mathbf{q}}, \nonumber
\end{eqnarray}
where the summation runs over all phonon modes $j$, the average is over all phonon wave vectors $\mathbf{q}$ in the surface Brillouin zone, $m_{a,b}$ are the masses of the respective atoms and $\mathbf{x}_{a,b}$ their equilibrium coordinates, $\omega_{\mathbf{q} j}$ are the phonon frequencies, $N_{\mathbf{q} j}\equiv[\exp(\beta \hbar \omega_{\mathbf{q} j})-1]^{-1}$ are the phonon occupancies, and $e_{\mathbf{q} j}^{a \alpha}$ are the displacements for atom $a$ in direction $\alpha$ associated with the $\mathbf{q},j$-phonon mode. Moreover, by combining Eqs.~\ref{eq:dD2},\ref{eq:uu} and averaging over $\mathbf{r}$, we can define an \emph{elastic scattering attenuation} spectral function,
\begin{eqnarray} \label{eq:spectral}
 g(\omega) &=&\frac{\hbar}{2  \sqrt{m_a m_b}} \sum_{a,b,\alpha,\beta} \left\langle\frac{\partial_{a\alpha} n(\mathbf{r})}{\partial_{\hat n} n(\mathbf{r})} \frac{\partial_{b\beta} n(\mathbf{r})}{\partial_{\hat n} n(\mathbf{r})}\right\rangle_{\mathbf{r}} \cdot\\ &&  \left\langle \sum_{j} \frac{1}{\omega_{\mathbf{q} j}}\hat e_{\mathbf{q} j}^{* a \alpha}\hat e_{\mathbf{q} j}^{b \beta} e^{i\mathbf{q}\cdot (\mathbf{x_b}-\mathbf{x_a})}\delta(\omega-\omega_{\mathbf{q} j})\right\rangle_{\mathbf{q}}, \nonumber
\end{eqnarray}
where, in the first line, we use the shorthand $n(\mathbf{r},D(\mathbf{r}))\rightarrow n(\mathbf{r})$. We now can closely approximate the Debye-Waller exponent appearing in $I=I_0 e^{-2W}$ by averaging $\delta D^2(\mathbf{r})$ across the surface. The final result is
\begin{equation}\label{eq:esa}
W \approx \Delta k_z^2\int d\omega \, g(\omega) \left(N(\omega,T)+\frac{1}{2}\right), 
\end{equation}
where $N(\omega,T)$ is the Bose occupancy factor.

\emph{Results and discussion.---}
Figure~\ref{fig:intensity}\,a compares our measurements \cite{thompson2024correlating,doi:10.1021/acs.jpcc.4c02430} with our \emph{ab initio} predictions from Eqs.~\ref{eq:pmS},\ref{eq:dD2},\ref{eq:uu} for changes in the specular reflection intensities over the range $\sim$\,500--2000~K. It is important to note that since the overall normalization $I(T=0)$ at low temperatures ($T\rightarrow 0$) is experimentally inaccessible, the experimental data estimates this value through linear extrapolation to $T=0$ \cite{thompson2024correlating,doi:10.1021/acs.jpcc.4c02430}. Additionally, to allow a direct comparison, we use the same extrapolation-normalization procedure to plot our theoretical predictions. The solid lines in the figure show the resulting theoretical predictions for each surface, using the respective average density contours of $n_{c,\text{metal}}=1.30$~nm$^{-1}$ and $n_{c, \text{oxide}}=0.66$~nm$^{-1}$ determined above. To estimate the uncertainty in our density-contour model, the dashed lines indicate prediction ranges that result from varying $n_c$ by $\pm 0.20$~nm$^{-1}$, according to the rms density variations along the turning-point surfaces determined earlier. The agreement between theory and experiment is remarkable, especially given that the predictions involve no adjustable parameters and that the match is equally strong for both the nearly flat, single-species Nb(100) surface and the highly corrugated, multi-species (3$\times$1)-O/Nb(100) surface.

With our theoretical approach confirmed via experiment, we can now confidently separate the contributions from various phonon modes by restricting the summations in Eq.~\ref{eq:uu}, thereby providing a microscopic understanding of a number of observed phenomena. For example, in the low-temperature regime, which we have not yet accessed within our own apparatus, there have been reports of nonlinear behavior in Debye-Waller exponents \cite{benedek2022role}. Linear behavior is expected above temperatures where most phonon modes are thermally accessible, and their occupations in Eq.~\ref{eq:uu} thus follow \( N_{\mathbf{q}j} \sim T \). Figure~\ref{fig:intensity}\,b shows our predicted total specular intensities at low temperatures for both the metallic and oxidized surfaces, including approximated values from Eq.~\ref{eq:esa}, a linear extrapolation and the spectral functions $g(\omega)$ defined in Eq.~\ref{eq:spectral}.

The energy scale for the phonon density of states has been matched with the temperature scale used in helium atom scattering (HAS) measurements of attenuation, so that the phonon energy $\hbar\omega$ corresponds to a temperature of 
$1.884\, k_B T$. This is the point where the slope of the Bose-Einstein occupancy function reaches 75\% of its classical equipartition value, indicating a shift to more non-linear behavior in the system’s response. Essentially, this phonon energy marks when quantum effects become strong enough to result in non-linear phonon behavior.  For both surfaces, as we would predict, the non-linear behavior of the Debye-Waller slope indeed becomes noticeable when our scaled temperature reaches the median phonon energy in $g(\omega)$ (vertical dashed lines), at around \( T \sim 80~\text{K} \) and \( T \sim 120~\text{K} \) for the metal and oxide respectively.

Beyond analyzing thermal attenuation by separating contributions from different phonon frequency ranges, we can also use our \emph{ab initio} approach to isolate contributions from individual phonon modes, based on their atomic displacement patterns. This allows us to distinguish the contributions of surface, resonant, and bulk modes, as well as different polarizations.
These effects have been considered in earlier studies \cite{mcmillan2022combined, benedek1994helium}, but there is limited direct knowledge of the quantitative contributions of individual surface modes to the Debye-Waller factors observed in specific systems.

\begin{figure}[!htb]
   \centering
   \includegraphics*[width=1\columnwidth]{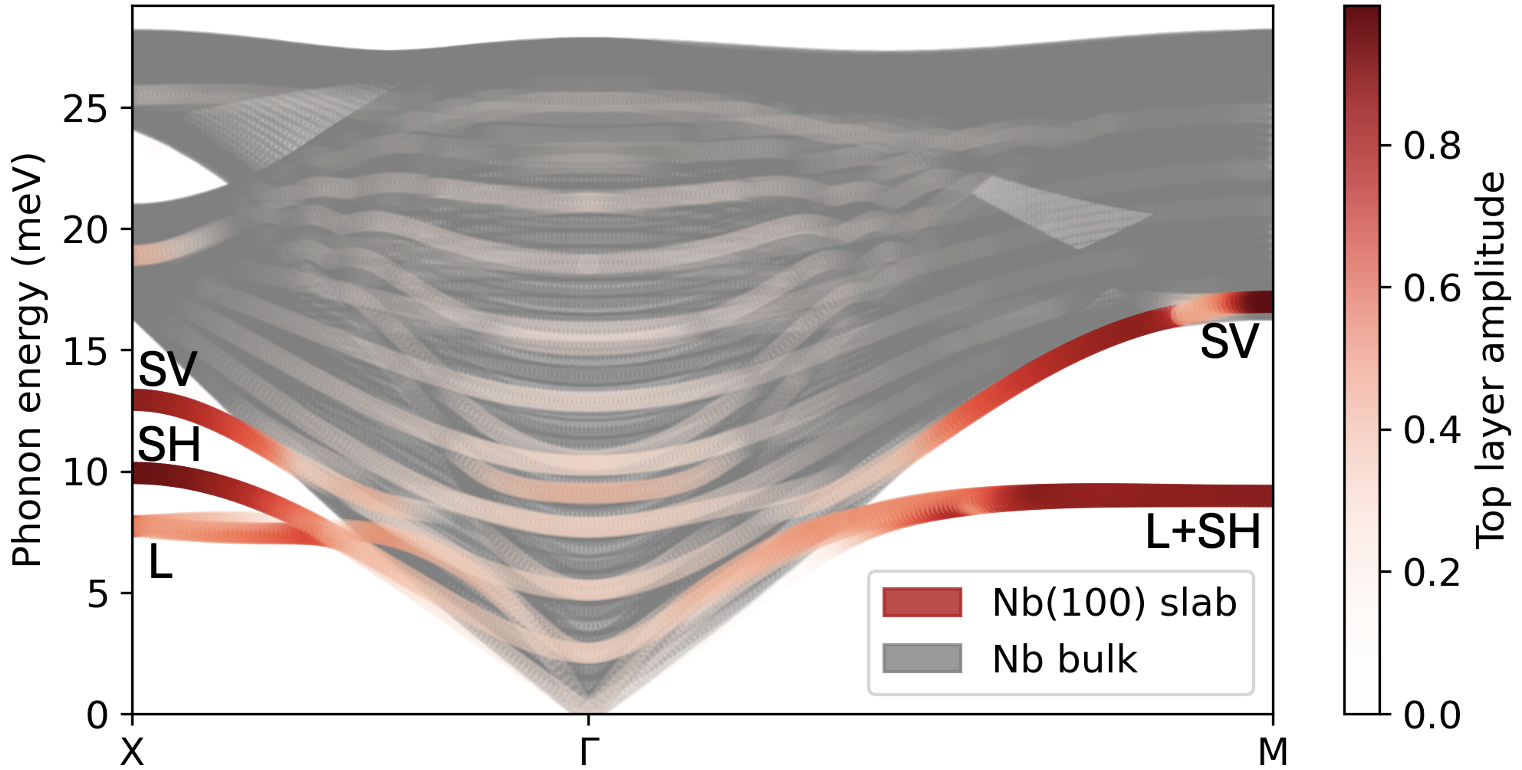}
   \caption{Phonon spectrum for Nb(100):  frequency versus wave vector, colored according to the fraction of mode amplitude associated with surface atoms (white$\rightarrow$red colormap), with bulk modes (grey) and labels for surface longitudinal (L), shear horizontal (SH), and shear vertical (SV) modes.}
   \label{fig:phonons}
\end{figure}

We begin by separating surface and resonant modes from bulk modes by defining the \emph{surface participation} of each mode as $\sigma_{\mathbf{q}j} = \sum_{a\alpha} \left|\hat e_{\mathbf{q} j}^{a \alpha}\right|^2$, where $a$ ranges over atoms on the surface and $\alpha$ over the three coordinate directions. Figure~\ref{fig:phonons} shows the resulting participation for the Nb surface phonon spectrum overlaid with the projected bulk phonon spectrum, clearly showing the separation into true surface states, resonant states in the energy range of the bulk modes, and true bulk modes. The first two lines of Table~\ref{tbl:surface} show the relative contributions of bulk and surface-active modes (including both resonant and true surface modes), as determined by including respective factors of $(1-\sigma_{\mathbf{q}j})$ and $\sigma_{\mathbf{q}j}$ in the sum in Eq.~\ref{eq:uu} and then using the result in Eqs.~\ref{eq:pmS},\ref{eq:dD2}. Interestingly, whereas we find the thermal attenuation in both systems to be dominated by surface-active modes, this is much more the case for the oxide surface ($\sim$\,6:1 versus $\sim$\,2:1). We attribute this to the strong surface activity associated with high-frequency modes involving the low-mass oxygen atoms.

To analyze the effects of surface activity in more detail, the third and fourth lines of Table~\ref{tbl:surface} decompose the total surface contributions into resonant and true surface contributions, defined by restricting the sum in Eq.~\ref{eq:uu} to modes whose phonon energies are or are not, respectively, within 0.1~meV of any bulk mode at the same wave vector. We find that the surface contributions for the metal surface are distributed roughly equally between resonant and true surface modes, whereas for the oxide surface, true surface modes contribute more than three times as much as the resonant modes and result in \emph{two-thirds of the total} thermal attenuation effect. Again, this is likely associated with the presence of low-mass oxygen atoms at the surface. Another intriguing difference between the metal and oxide surfaces becomes apparent when we analyze the surface contributions into longitudinal, shear-horizontal, and shear-vertical components, as obtained by restricting the sum over $\alpha$ in Eq.~\ref{eq:uu} to include only the respective components of the displacement (final three lines of Table~\ref{tbl:surface}). The  metal surface behaves as expected, with the surface contributions strongly dominated by the shear-vertical component. The oxide surface, however, shows roughly equal contributions from all three polarizations. We attribute this to the highly corrugated nature of the surface, where horizontal motions of the surface atoms can significantly impact the location of the hard corrugated surface, particularly at the locations of highest slope descending into the trenches evident in Fig.~\ref{fig:top}~c.

\begin{table}
\begin{tabular}{|c||c|c|} \hline
DW slope contribution & Nb(100)  & (3$\times 1$)-O/Nb(100) \\  \hline \hline

\hline

Bulk    & 35.6\% & 13.4\% \\ \hline
Surface & 64.4\% & 86.6\% \\ \hline \hline

Resonant Surface    & 30.2\% & 20.3\% \\ \hline
True Surface & 34.2\% & 66.3\% \\ \hline \hline

Longitudinal    & 7.4\% & 26.8\% \\ \hline 
Shear-horizontal    & 4.6\% & 25.6\% \\ \hline
Shear-vertical    & 52.4\% & 34.2\% \\ \hline

\end{tabular}
\caption{Fractional contributions to the net Debye-Waller (DW) slope for Nb(100) and (3$\times 1$)-O/Nb(100) from bulk and surface modes; from resonant and true surface modes; and from surface longitudinal, surface shear-horizontal, and surface shear-vertical modes.
}
\label{tbl:surface}
\end{table}

\emph{Conclusions. ---} This study presents a fully \textit{ab initio} method for predicting thermal attenuation in elastic helium atom scattering amplitudes, which we validate through strong agreement with our measurements on both the smooth, single-species Nb(100) surface and the highly corrugated, multi-species (3$\times 1$)-O/Nb(100) oxide surface. Our theoretical method directly provides \emph{ab initio} evidence of the phonon quantization mechanism, which underlies recent observations of nonlinear behavior in Debye-Waller exponents at low temperatures\cite{benedek2022role}. Additionally, our method offers direct quantitative insights into the relative contributions of bulk, surface, and resonant modes, as well as different surface polarizations, revealing key contrasts between single-species smooth surfaces and multi-species corrugated surfaces. These findings improve our understanding of surface dynamics and electron-phonon coupling, paving the way for further experimental and theoretical investigations into surface superconductivity.

\emph{Acknowledgements. ---} This work was supported by the US National Science Foundation under award PHY-1549132, the Center for Bright Beams.

\bibliographystyle{myunsrt}
\bibliography{PRL}
\end{document}